# Magnonic waveguide based on exchange-spring magnetic structure

Lixiang Wang, Leisen Gao, Lichuan Jin, Yulong Liao, Tianlong Wen, Xiaoli Tang, Huaiwu Zhang, Zhiyong Zhong[1]

State Key Laboratory of Electronic Thin Films and Integrated Devices, University of Electronic Science and Technology of China, 610054, Chengdu, China

**Abstract**

We propose to use a soft/hard exchange-spring coupling bilayer magnetic structure to introduce a narrow channel for spin-wave propagation. We show by micromagnetic simulations that broad-band Damon-Eshbach geometry spin waves can be strongly localized into the channel and propagate effectively with a proper high group velocity. The beamwidth of the bound mode spin waves is almost independent from the frequency and is smaller than 24nm. For a low-frequency excitation, we further investigate the appearance of two other spin beams in the lateral of the channel. In contrast to a domain wall, the channel formed by exchange-spring coupling can be easier to realize in experimental scenarios and holds stronger immunity to surroundings. This work is expected to open new possibilities for energy-efficient spin-wave guiding as well as to help shape the field of beam magnonics.

**Introduction**

Magnonics is an emerging technology for low-power signal transmission and data processing based on spin waves (magnons) propagating in magnetic materials [1-4]. Nowadays such magnon-based computing concept is discussed and undergoes benchmarking in the framework of beyond-CMOS strategies [5], due to its nanometer wavelengths and Joule-heat-free transfer of spin information over macroscopic distances [6-8]. In the context of the magnonics [9-14], where the control of spin waves is sought as a practical means of transmitting and processing information (in the same vein as the control of light in photonics), the capacity for energy-efficient propagation of spin waves is essential for any form of circuit design, and is crucial for wave processing schemes that rely on spin wave interferences [13-17]. In the last few years, magnonic crystals have been widely studied [10, 18-24]. One of these artificial crystals is periodically patterned at the nanoscale to promise a degree of control of spin waves when transmitting through magnetic materials. However, limited by the current nanotechnology, a considerable challenge for this kind of magnonic crystals is to fabricate perfect structures precisely and controllably at the submicron or nanometer scale. Recently, individual domain walls were functionalized as nanoscale magnonic conduits that allowed for a rewritable nanocircuitry in several papers [25-30]. A magnetic domain wall acts as a confining potential well, spin waves sent along the wall can be strongly localized into its center but propagate freely in the direction parallel to the wall without extra fields. Unfortunately, these specific domain walls are usually difficult to realize in experimental scenarios. Besides, fragility is also a considerable problem, since a domain wall structure can be easily invalidated by stray fields from surroundings. Here, we will provide a new possible way to overcome these challenges. We present a paradigm for spin-wave propagation that relies on a domain-wall-like magnetic channel as magnonic waveguides. We emphasize that the channel is naturally induced on a soft/hard bilayer magnetic structure via exchange-spring coupling interaction, thereby can be easily realized in laboratory and possesses strong immunity to disturbance from outside. We show through micromagnetic simulations that a Damon-Eshbach (DE) propagation geometry sent to the channel can be strongly confined in a narrow area with a broad frequency band. By targeting the bonded modes, we focus on the potential of using such magnetic structures as channel inductors for spin-wave guiding to open new perspectives for efficient spin-wave propagation towards magnonic nanocircuitry.

---

[1]Author to whom correspondence should be addressed. Electronic mail: zzy@uestc.edu.cn

**Micromagnetic modeling**

Fig. 1(a) sketches the geometry sizes of the model we used. The model includes a soft magnetic layer (SL) that is exchange coupled to a hard magnetic layer (HL) at the bottom surface of the SL [31]. The thicknesses of two layers are $t_{SL} = 12nm$ and $t_{HL} = 8nm$, respectively. The HL magnetization is oriented in the plane-normal direction to model stripe domains of a width 100nm along the x direction. Periodic boundary conditions in the y direction are used to model an infinite array of stripe domains. The SL magnetization is initially randomized with an in-plane random vector field, and is then allowed to equilibrate at zero applied fields. The whole simulation process contains two steps, which are stabilizing the system and propagating spin waves successively. In the second step, an out-of-plane field pulse was used to locally excite spin dynamics at the channel's position at a distance of 40nm from the left edge (green area in Fig.1b). The OOMMF micromagnetic code [32] was used to carry out simulations by solving the Landau-Lifshitz-Gilbert equation.

Values of the magnetic saturation of the SL, $M_S^S = 8.0 \times 10^5 \, A/m$, the exchange stiffness $A^S = 13 \times 10^{-12} \, J/m$ and the anisotropy, $K^S = 0 \, J/m^3$ which were taken from NiFe, were used in our numerical simulations[33]. Similarly, standard values of magnetic saturation of [Co/Pd]$_5$, $M_S^H = 3.6 \times 10^5 \, A/m$, the perpendicular anisotropy, $K^H = 6.3 \times 10^5 \, J/m^3$, and the exchange stiffness, $A^H = 6.0 \times 10^{-12} \, J/m^3$ [33], are taken to model the HL. The exchange between the SL and HL was modeled with an intermediate value $A^{SH} = 9.5 \times 10^{-12} \, J/m^3$. The model size in the x direction was set to 3μm to minimize boundary effects and provide a sustainable spin-wave propagation. The damping parameter was set at $\alpha = 0.5$ in the first step to lead to rapid convergence to get grounded magnetization state but $\alpha = 0.01$ (was taken from NiFe) in the second step to support spin waves for a long-distance propagation.

**Results and discussions**

Here, we focus on the possibility of bringing about an expected magnonic waveguide based on soft/hard magnetic structure by exchange-spring coupling. Fig. 1b shows the remnant magnetization configuration in the top surface of the SL after the stabilization of the system. The orientation of arrows represents the magnetization vectors, with red-green color codes the normalized magnetization component $m_y$ along the width direction of the model. In the Fig. 1b, one observes a noticeable strip area (indicated by the dashed box) at $y = 200\text{nm}$ along the x axis, which separates two almost opposite domain patterns. The area is characterized by inside magnetic moments that are strictly oriented to the y direction. It is particularly similar to a 180° Néel wall, but the magnetization within it is in a better order. Besides, we emphasize that such magnets have been found to display characteristic structure with enhanced remnant magnetization [33-35], thus the magnetization can exhibit stronger immune to stray fields compared with a domain wall. In the following, we utilize such specific strip magnetization as a narrow channel for spin-wave propagation. Due to the magnetic moments within the channel are perfectly parallel to the y axis, no external bias fields were needed in our simulations to either induce the channel or propagate the spin waves.

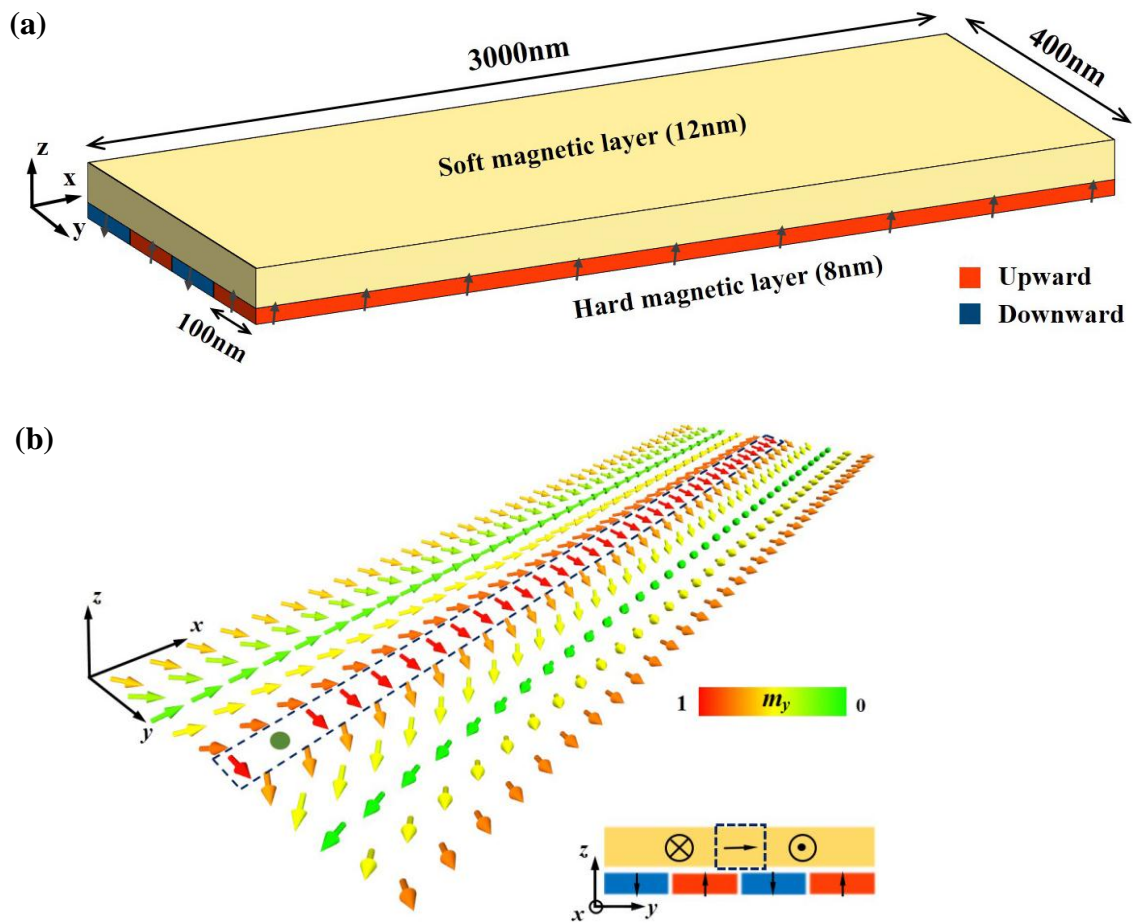

**Figure 1 (a)** Schematic illustration of the magnetic structure used in simulations. Red-blue areas in the HL denote the magnetic moments directing up and down, respectively. **(b)** Stabilized magnetization configuration on the top surface of the SL, with the discussed channel (indicated by the dashed box) obtained from the OOMMF simulation. The color of vectors codes $m_y$ values as indicated by the color bar. Spin waves were locally excited at the position indicated by the green dot. The lower right corner sketches the cross section in the y-z plane.

To verify the channeling effect, single-frequency spin waves propagating along the channel were investigated first. An oscillating magnetic field of $h_z = H_0 \sin(\omega t)$ with $H_0 = 100 mT$, to render higher phase resolution between neighboring simulation cells, was applied to the marked position in Fig. 1b to generate spin dynamics. The $M_z$ component of the magnetization was plotted at t = 2.0ns after the excitation to ensure the system reaching a steady state. Fig. 2a shows the snapshots of spin wave distribution at various frequencies. For the lower excitation frequencies of 5 and 12GHz, spin waves exist only inside the channel as expected for a bound mode, justifying that the channel formed by the studied structure can be indeed served as a waveguide for spin-wave propagation. Similar to a domain wall, the channeling effect can be understood as following [28]. The perfectly ordered magnetization within the channel separates two opposite domains, giving rise to opposite magnetic volume charges on the two sides of the channel. These charges generate a strong magnetostatic field $H_{demag}$ antiparallel to the magnetization within the channel, resulting in a locally decreased effective field which forms a magnetic potential well for bond modes. As presented in Fig. 2a, the well in the internal field is so deep that spin waves can be completely localized into the channel with much pronounced strength. It is noteworthy that the magnetization of the channel is oriented along the y direction and thus perpendicular to the propagation direction of the spin waves, which forms the DE propagation geometry. It is well known that DE spin waves can be easily generated and controlled, have positive dispersion and, more importantly, high group velocity [36]. The slight decrease in amplitude for long-distance propagation is related to the introduction of Gilbert damping α in the simulations (here $\alpha = 0.01$ was used). In contrast, once the excitation frequency reaches 25GHz, spin waves are spread throughout the domains at both sides of the channel, the channeling effect disappears. The observation can be analogized to the spin waves propagating in a domain wall [28-30, 37]: spin dynamics with a high frequency lie in the allowed band instead of the energy gap of the extend spin-wave modes and, thus are mixed with the extend modes, resulting in the loss of the channeling effect.

Here, we focus on the spin-wave mode localized inside the channel. The beamwidth of the well-mode spin wave for various frequencies, which is defined as the full width at half maximum (FWHM) of the well in $M_z$, was studied below. As shown in Fig. 2b, all the calculated data of the beamwidth is smaller than 24nm and is almost irrelevant to the excitation frequency, which is in good agreement with what is expected from a waveguide width at nanoscale. Thanks to the perfect confining effect of the channel, spin waves are strongly squeezed to a narrow alley, resulting in the weakness of magnetization precession at the lateral edges [38]. In other words, spin waves propagating along the channel are insensitive to the edge, as the light traveling in a fiber. Such optic-fiber-like waveguide should have great potential for magnonic transmission [25-27], since the negligible boundary scattering experienced by the channeled spin waves will lead to an increased propagating length.

In addition to spatial and spectrum characteristics, we also shed light on the spin-wave dispersion to get a deeper insight into the well-confined mode. A symmetric *sinc* field pulse of $h_z(t) = H_0 \frac{\sin(2\pi f_c t)}{2\pi f_c t}$, ranging from 0 to 25GHz, was used to the excite spin waves within a frequency range. Two-dimensional fast Fourier transform (2D FFT) of $M_z$ is performed along the propagating direction to plot the dispersion curve [39]. Fig. 2c shows the resulting positive dispersion that enables the transport of information via spin waves propagating within the channel. Parabolic dispersion relation with large wave vectors (corresponding to small wavelength) suggests that the bond mode is mainly dominated by exchange energy. While the wide frequency range of the dispersion curve illustrates that the critical frequencies, within which spin waves are permitted to be confined in the channel, can be close to 0 and 22GHz, respectively. The observation of well-defined wave vectors along the propagation path is a crucial precondition for numerous applications that rely on the interference of spin waves and highlights the potential of the channel in magnonic circuits for data processing.

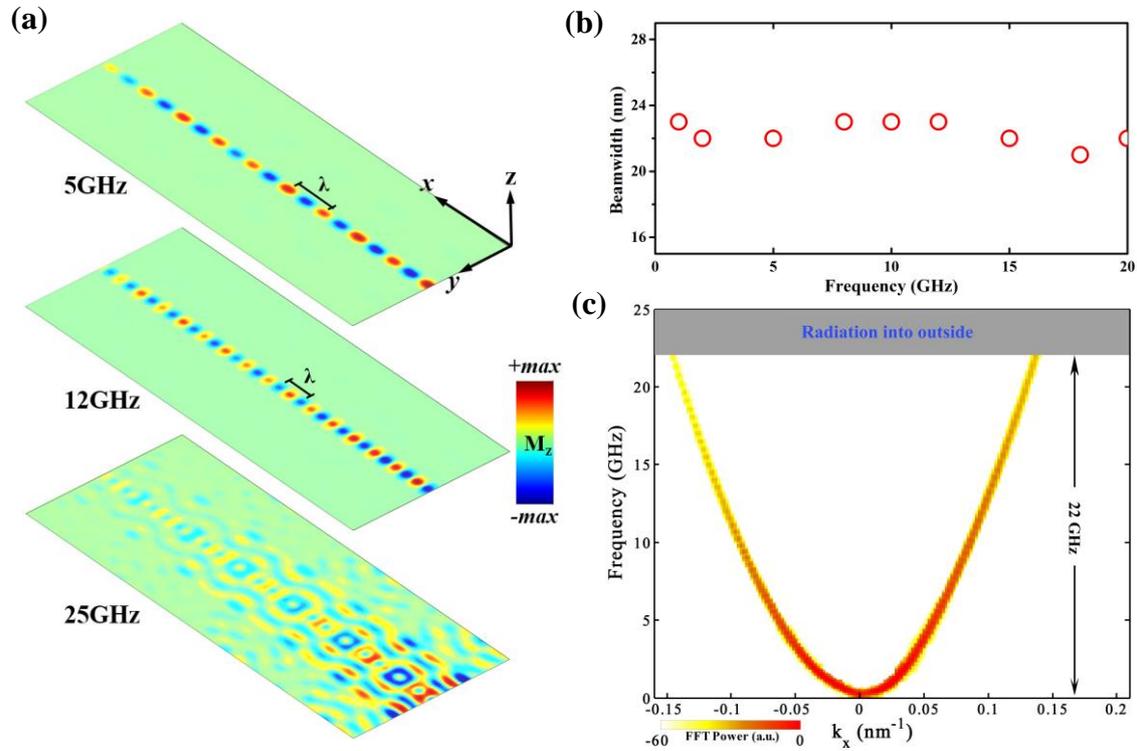

**Figure 2** (a) Spin-wave propagation patterns at $t = 2.0$ns for labeled frequencies. The color codes the out-of-plane component of the magnetization $M_z$ as indicated by the color bar. (b) The beamwidth as a function of frequency. The beamwidth is defined as the FWHM of the SW amplitude distribution over the waveguide width. (c) Dispersion relation of the well-confined mode calculated by 2D FFT.

We further investigate the group velocity of the bound mode. An illustration of wave packet propagating along the channel at labeled times is shown in Fig. 3a. The wave packets are generated with a field pulse that comprises a sine wave oscillation at the frequency $f = 5$GHz. The temporal evolution of the normalized $m_z$ component of the wave packet is shown for three instants after the application of the pulse field. By relating the evolution time $\Delta t$ to the propagated distance $\Delta x$ we can extract the wave packet velocity, which $v_g = \Delta x/\Delta t$ is approximately 1440m/s for the given frequency. In fact, the upward curve plotted in Fig. 3b reflects a positive correlation between the group velocity and frequency, which can be derived from the spin-wave dispersion relation by $v_g = \partial \omega(k)/\partial k$. The group velocity can exceed 1km/s at 1GHz, while over 8GHz the velocity exceeds 1.5km/s. Large value of the propagation velocity brings competitiveness to the well-defined mode for computing technology, since the velocity determines the speed of calculation [7].

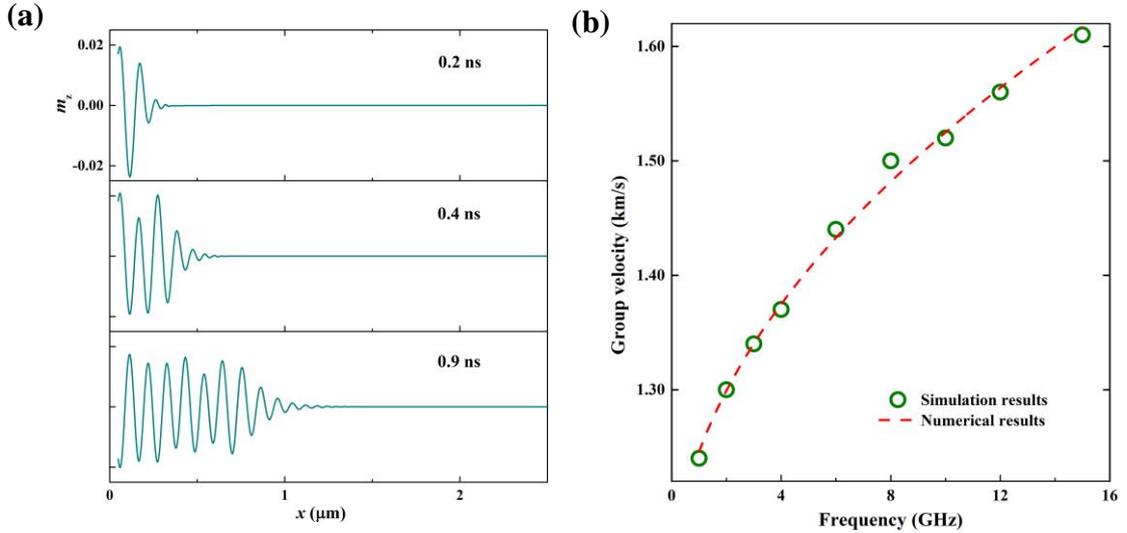

**Figure 3** **(a)** Illustration of wave packet propagating along the channel at various time. **(b)** The group velocity as a function of frequency. Green circles are simulation results obtained from OOMMF, while red dashed line is corresponding numerical results calculated by $v_g = \partial \omega(k)/\partial k$.

When the excitation frequency applied in the channel was reduced to a very low value, for example 2GHz, an attractive observation emerged: two other spin-wave beams parallel to the channel are presented at $y = 100$nm and $300$nm respectively except for the discussed one (see Fig. 4a for details). Clearly, the appearance of them suggests the existence of other propagating channels at corresponding positions. To get a deeper insight of these channels, we extracted the x component of the internal field $H_{eff}^x$ across the y direction and Fig. 4b (the bottom panel) shows the result.

Two equal-height but antisymmetric peaks on the curve prove that channels for spin-wave propagation indeed exist there and the potential wells are equal-deep but their magnetization is opposite (see the top panel in Fig. 4b). Below we label this two channels as lateral channels (LCs) while the previous one as middle channel (MC) to distinguish from each other. Note that the magnetic moments in the LCs are oriented to the $x(-x)$ axis and thus parallel to the propagation direction of spin waves, which forms the backward-volume-wave propagation [36]. It is quite different from the MC mode.

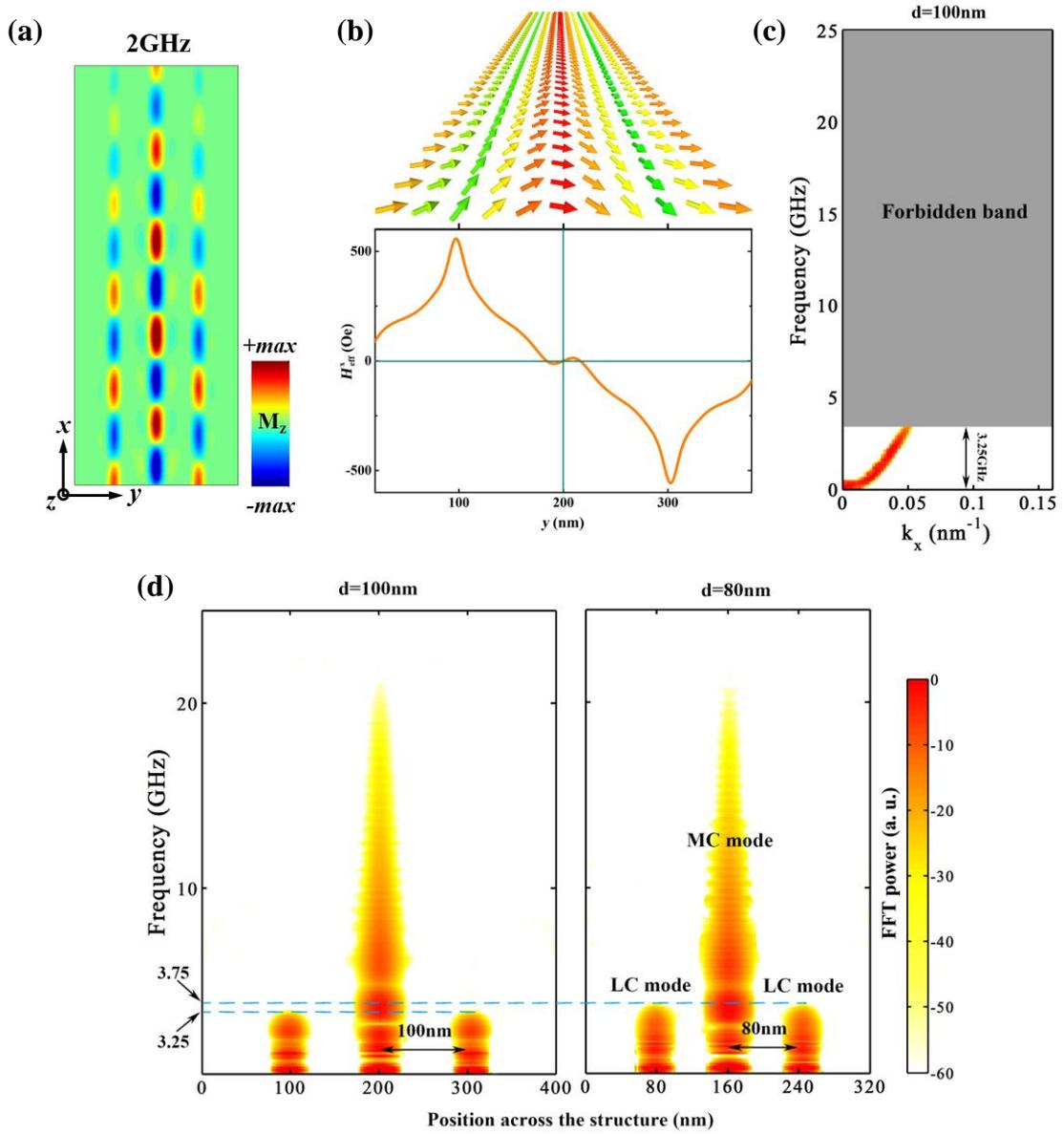

**Figure 4** (a) Spin-wave propagation pattern at t = 2.0ns for f = 2GHz. (b) The x component distribution of the internal field over the width (the bottom panel) and the 3D view of the spin configuration (the top panel). (c) The dispersion curve of the LC modes for w=400nm. (d) Energy spectrum of spin waves across the y direction for w = 400 and 320nm.

Again, we should be noted that the LCs modes appear only when the excitation frequency is lower, they will vanish at a high excitation frequency (Fig. 2a), as indicated by their dispersion curve in Fig. 4c. We propose to give the explanation as following. It is well known that spin dynamics are dominated by dipole or exchange energy depending on the wavelength. For the studied mode with a low-frequency excitation, which corresponding to a large wavelength (Fig. 2c), spin waves in the MCs are dominated by dipole-dipole interaction. The dynamic dipole energy characters a wide functioning range which can cross the barriers between the channels, powering the spin precession in the LCs. However, when the frequency is at a high value (corresponding to a small wavelength), exchange interaction charactering narrow functioning range will take place of the dominating role of dipole-dipole interaction, resulting in the loss of driving force on the LCs, hence the LC mode disappears. To verify the words above, we decreased the distance between the channels from d = 100nm to 80nm and the cut-off frequency of the LCs modes increased as expected (shown in Fig. 4d). It can be understood that shorter distance between the adjacent channels allows narrower-range interaction, which characterizes higher frequency, to drive the LCs modes, accordingly resulting in the higher cut-off frequency. The appearance of such LC modes is a fantastic observation and might provide an unprecedented possibility for spin-wave filter design.

**Conclusion**

In summary, we have demonstrated that a soft/hard exchange-spring coupling bilayer magnets with certain magnetization can cause a deep potential well for spin-wave channeling. Because the channel is primarily governed by intrinsic interplay between two layers, it is less sensitive to experimental scenarios. Spin waves with a DE propagation geometry exhibit a well-defined wave vector along the channel, enabling data transport and processing using wave properties. The beamwidth of the bound mode is smaller than 24nm and is almost independent from frequency, which can avoid the boundary scattering [38] caused by edge irregularity and the extra dispersion [40]. In addition, a relative high group velocity exceeding 1km/s promises the channeled mode a candidate for computing technology. Finally, we have addressed and qualitatively verified the appearance of LC modes, which provides a new train of thought for spin-wave filter design. These observations pave the way for the realization of nano-sized, energy-efficient, reconfigurable magnonic circuits.

**Acknowledgment**

This paper is supported by the National Key Research and Development Plan (No. 2016YFA0300801); the National Natural Science Foundation of China under grant Nos. 61734002, 61571079 and 51702042, and the Sichuan Science and Technology Support Project (Nos. 2016GZ0250 and 2017JY0002).